\documentclass[a4paper]{jpconf}
\usepackage{graphicx}
\usepackage{booktabs}
\usepackage{lmodern}
\usepackage[T1]{fontenc}

\usepackage{amsmath}
\usepackage{amssymb} 
\usepackage{cite} 

\usepackage[hidelinks]{hyperref}

\hypersetup{
    pdfauthor={Nathan Clisby and Dac Thanh Chuong Ho},
    pdftitle={Off-lattice and parallel implementations of the pivot algorithm}
    }

\usepackage{tikz}
\usetikzlibrary{shapes}
\usetikzlibrary{trees}
\usetikzlibrary{patterns}
\usetikzlibrary{positioning}
\usetikzlibrary{decorations.pathmorphing}
\tikzset{cross/.style={cross out, draw=black, minimum
size=2*(#1-\pgflinewidth), inner sep=0pt, outer sep=0pt},
cross/.default={2pt}}

\newcommand{\Z}{{\mathbb Z}}
\newcommand{\R}{{\mathbb R}}

\newcommand\tstrut{\rule{0pt}{2.4ex}}
\newcommand\bstrut{\rule[-1.0ex]{0pt}{0pt}}

\usepackage{float}

\begin{document}

\title{Off-lattice and parallel implementations of the pivot algorithm}
	
\author{Nathan Clisby and Dac Thanh Chuong Ho} 

\address{Department of Mathematics, Swinburne University of
Technology, P.O. Box 218, Hawthorn, Victoria 3122, Australia}

\ead{nclisby@swin.edu.au}

\begin{abstract}
The pivot algorithm is the most efficient known method for sampling
polymer configurations for self-avoiding walks and related models.  Here
we introduce two recent improvements to an efficient binary tree implementation
of the pivot algorithm: an extension to an off-lattice model, and a 
parallel implementation.
\end{abstract}

\section{Introduction}
\label{sec:intro}

Self-avoiding walks are non-intersecting paths on lattices such as the
two-dimensional square lattice or the three-dimensional simple cubic lattice. 
Due to universality, they exactly
capture the essential physics of the excluded-volume effect for polymers
in the good-solvent limit, and as such can be used to study features
such as the value of the Flory exponent $\nu$ which relates the
geometric size of a walk to the number of monomers in the chain.

The pivot algorithm is the most efficient known method for sampling
self-avoiding walks of fixed length. It is a Markov chain Monte Carlo
method, which was invented by Lal~\cite{Lal1969MonteCarlocomputer}, but
first studied in depth by Madras and
Sokal~\cite{Madras1988PivotAlgorithmHighly}, who also invented an
efficient hash table implementation.  Recent improvements to the
implementation of the pivot
algorithm~\cite{Kennedy2002fasterimplementationpivot,Clisby2010AccurateEstimateCritical,Clisby2010Efficientimplementationpivot}
have dramatically improved computational efficiency to the point where
it is possible to rapidly sample polymer configurations with up to 1
billion monomers~\cite{Clisby2018MonteCarlo4dSAWs}.

In this paper, we will describe two recent improvements in algorithms to
sample self-avoiding walks, focusing in particular on the pivot
algorithm.
In Sec.~\ref{sec:offlattice} we describe an off-lattice implementation
of the SAW-tree data
structure~\cite{Clisby2010Efficientimplementationpivot}.
In Sec.~\ref{sec:parallel} we describe a parallel implementation of the
pivot algorithm which improves the sampling rate for very long walks.
Finally, we have a brief discussion about prospects for further progress
and conclude in Sec.~\ref{sec:conclusion}.

\section{Off-lattice implementation}
\label{sec:offlattice}

The SAW-tree data
structure~\cite{Clisby2010Efficientimplementationpivot} is a binary tree
that encodes information about the self-avoiding in an efficient way in
nodes in the tree. In
particular, the leaves of the tree consist of individual monomers, while
the internal nodes store aggregate information about all of the
monomers that are below that node within the tree, as well as
``symmetry'' information which encodes transformations that must be
applied to sub-walks before they are concatenated together.
The aggregate information that must be stored includes information about
the extent of the sub-walk in the form of a ``bounding volume'', which is
taken to be a rectangle for square-lattice walks, and a rectangular
prism for simple-cubic-lattice walks.
For lattice self-avoiding walks, the symmetry elements are rotations and
reflections that leave the lattice invariant.
See~\cite{Clisby2010Efficientimplementationpivot} for a full description
of the implementation.

Although lattice self-avoiding walks capture the universal behaviour of
polymers in the good-solvent limit, there are strong arguments for why
off-lattice models of polymers may have advantages under certain
circumstances. Firstly, they provide an opportunity to empirically model
more realistic interactions for polymers, and thus to reproduce not only
universal features but also make precise experimental predictions.
Secondly, under some circumstances it may be the case that the effect of
the lattice may have a non-negligible effect, for example when trying to
understand the nature of the globule transition it may be the case that
the restriction to the lattice significantly influences the nature of
the transition. Finally, while lattices have discrete symmetry groups,
the symmetry group corresponding to reflections and rotations of $\R^d$
is the continuous orthogonal group $O(d)$. This continuous group 
allows for more freedom for performing pivot moves, and it is
conceivable  that this additional freedom
may enhance sampling efficiency under some circumstances.

\begin{figure}[htb]
\begin{center}
  \begin{tikzpicture}[scale=0.30, thick, draw]
\fill ( 0.000000000000000e+00, 0.000000000000000e+00) circle (0.5);
\fill ( 9.786616592965383e-01, 2.054783604736682e-01) circle (0.5);
\fill ( 7.138507999451933e-01,-7.588220122160878e-01) circle (0.5);
\fill ( 1.652016368820576e+00,-1.105008905926305e+00) circle (0.5);
\fill ( 2.646246004809292e+00,-9.977362117224877e-01) circle (0.5);
\fill ( 2.489134802852920e+00,-1.985317130037450e+00) circle (0.5);
\fill ( 1.586240769818234e+00,-2.415180323519237e+00) circle (0.5);
\fill ( 1.247320086895478e+00,-3.355995273808173e+00) circle (0.5);
\fill ( 1.160688733842073e+00,-4.352235710985934e+00) circle (0.5);
\fill ( 7.015135092890352e-01,-5.240581428123505e+00) circle (0.5);
\fill ( 1.278387485331991e-01,-6.059664616128078e+00) circle (0.5);
\fill ( 9.829325913381517e-01,-6.578137874837782e+00) circle (0.5);
\fill ( 7.081814926340417e-01,-7.539653258885735e+00) circle (0.5);
\fill ( 1.695731958738442e+00,-7.696955760127714e+00) circle (0.5);
\fill ( 1.878679392456082e+00,-8.680078456690117e+00) circle (0.5);
\fill ( 1.728991642820777e+00,-9.668811776452891e+00) circle (0.5);
\fill ( 1.113220493093022e+00,-1.045673683073341e+01) circle (0.5);
\fill ( 3.795115065809678e-01,-9.777272980987558e+00) circle (0.5);
\fill (-2.104289417354798e-01,-1.058471973925722e+01) circle (0.5);
\fill (-5.009594578546239e-02,-1.157178272124454e+01) circle (0.5);
\fill (-1.042960620081759e+00,-1.145253617607652e+01) circle (0.5);
\fill (-1.745024251321964e+00,-1.074042195933189e+01) circle (0.5);
\fill (-2.285333890964720e+00,-1.158188823513212e+01) circle (0.5);
\fill (-3.279517601945027e+00,-1.168958572266747e+01) circle (0.5);
\fill (-3.014664949667944e+00,-1.072529682789485e+01) circle (0.5);
\fill (-3.886513092322947e+00,-1.023552053595228e+01) circle (0.5);
\fill (-4.882985538957040e+00,-1.015159995981588e+01) circle (0.5);
\fill (-5.158386379043892e+00,-1.111292944406124e+01) circle (0.5);
\fill (-6.156662931653451e+00,-1.105424448423360e+01) circle (0.5);
\fill (-6.375662239105125e+00,-1.202996949843365e+01) circle (0.5);
\fill (-7.032752317546095e+00,-1.278378156308533e+01) circle (0.5);
\fill (-6.934322713325797e+00,-1.377892557930260e+01) circle (0.5);
\fill (-6.990313442431963e+00,-1.477735686800211e+01) circle (0.5);
\fill (-7.486950105403038e+00,-1.564531540675437e+01) circle (0.5);
\fill (-7.455465187445388e+00,-1.664481963382957e+01) circle (0.5);
\fill (-7.664058957800536e+00,-1.762282200548775e+01) circle (0.5);
\fill (-8.591767582783518e+00,-1.799612711813984e+01) circle (0.5);
\fill (-9.548096314174508e+00,-1.770383384501146e+01) circle (0.5);
\fill (-1.042656985674248e+01,-1.722604284990159e+01) circle (0.5);
\fill (-1.106496930040173e+01,-1.645633761389526e+01) circle (0.5);
\fill (-1.171513371270300e+01,-1.569654406497756e+01) circle (0.5);
\fill (-1.259657486866839e+01,-1.522425020669611e+01) circle (0.5);
\fill (-1.283225528518851e+01,-1.619608082016170e+01) circle (0.5);
\fill (-1.382091634554016e+01,-1.604591655096177e+01) circle (0.5);
\fill (-1.455716917651184e+01,-1.672262311136013e+01) circle (0.5);
\fill (-1.523147064610889e+01,-1.746107929022200e+01) circle (0.5);
\fill (-1.621509248047241e+01,-1.764132381014795e+01) circle (0.5);
\fill (-1.721480257620690e+01,-1.761724634170690e+01) circle (0.5);
\fill (-1.741638639517049e+01,-1.859671760924467e+01) circle (0.5);
\fill (-1.833285653853490e+01,-1.819661452712505e+01) circle (0.5);
  \end{tikzpicture}
  \end{center}
  \caption{ Typical bead-necklace configuration with 50 monomers in $\R^2$.\label{fig:ths}}
\end{figure}
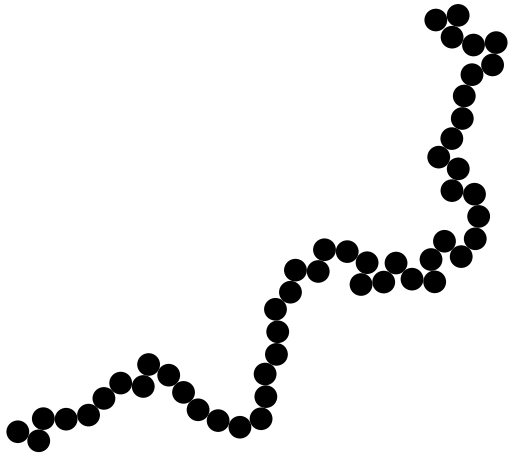

We implement the SAW-tree for the bead-necklace, or tangent-hard-sphere,
model, which consists of a fully flexible chain of hard spheres that
just touch. A typical configuration for this model in $\R^2$ is shown in
Fig.~\ref{fig:ths}.

We will now describe the key features of our implementation, and will
present evidence that the off-lattice SAW-tree implementation of the pivot
algorithm has $O(\log N)$ performance in line with the performance of
the original lattice SAW-tree implementation. The description will not
be self-contained, and the interested reader is referred
to~\cite{Clisby2010Efficientimplementationpivot} for relevant details.

The orthogonal group $O(2)$ is used as the symmetry group for $\R^2$,
and similarly $O(3)$ is used for $\R^3$.
The orthogonal group includes rotations as the subgroups $SO(2)$ and
$SO(3)$ respectively, but also includes reflection moves.

Symmetry group elements are sampled uniformly at random so as to
preserve the Haar
measure~\cite{Stewart1980EfficientGenerationOfRandomOrthogonalMatrices}
on the group. This automatically ensures that the Markov chain
satisfies the detailed-balance condition, and so must be sampling
configurations with uniform weights.

As for ergodicity, we feel that it is extremely likely that the
algorithm is ergodic. For lattice models the pivot algorithm has been
proved to be ergodic; this was first done for 
$\Z^2$ and $\Z^3$ in the seminal paper of Madras and
Sokal~\cite{Madras1988PivotAlgorithmHighly}. Interestingly, inclusion of reflections
seem to be necessary for ergodicity for lattice models.
In the continuum, it is our view that the additional freedom afforded
as compared to the lattice should mean that pivot algorithm is ergodic
in this case, too. We do not have sufficient insight into the problem to
know whether the extra freedom would allow one to have an ergodic
algorithm with only rotations
(and not reflections).
Some theoretical work has been done previously on the ergodicity of 
pivot moves for continuous
models~\cite{Plunkett2016OffLatticeSAWPivotAlgorithmVariant}, 
but this is not directly relevant here as the proof relied on
double-pivot moves.

The key decision for the SAW-tree implementation for the bead-necklace
model is the choice of \emph{bounding volume} to be used. The bounding
volume
is a shape which is stored in nodes in the SAW-tree, such that it is
guaranteed that the entire sub-chain which is represented by the node is
completely contained within the bounding volume.
The use of a bounding volume is necessary for the rapid detection of
self-intersections when a pivot move is attempted.

The natural choice of
the bounding volume for
$\Z^2$ is the rectangle, and for $\Z^3$ the natural choice is the
rectangular prism. This is because these shapes snugly fit the
sub-chains
that they contain (in the sense that the sub-chains must touch each boundary
or face of the shape), and the shapes are preserved under lattice symmetry
operations.

The natural shape for the bounding volume for the bead-necklace model
for $\R^2$ would seem to be the circle, and similarly for $\R^3$ the
natural choice would be the sphere. This is because these are the only
shapes that are invariant under the action of $O(2)$ and $O(3)$
respectively. 

\begin{figure}[htb]
\begin{center}
  \begin{tikzpicture}[scale=0.65, thick, draw]
       \begin{scope}[xshift=0.0cm,yshift=0.0cm]
       \draw[thick,decorate,decoration={snake,amplitude=.3mm,segment length=2mm,pre length=1.0mm, post length=1.0mm}] (0.4,0.15) -- (-0.61885,1.90211) ; 
       \draw[thick,decorate,decoration={snake,amplitude=.3mm,segment length=2mm,pre length=1.0mm, post length=1.0mm}] (-0.61885,1.90211) -- (-1.2,0.2);
       \draw[thick,decorate,decoration={snake,amplitude=.3mm,segment length=2mm,pre length=1.0mm, post length=1.0mm}] (-1.2,0.2) -- (-0.85156,-1.809654) ;
       \draw[thick,decorate,decoration={snake,amplitude=.3mm,segment length=2mm,pre length=1.0mm, post length=1.0mm}] (-0.85156,-1.809654) -- (0.7,-1.6);
       \draw[thick,decorate,decoration={snake,amplitude=.3mm,segment length=2mm,pre length=1.0mm, post length=1.0mm}] (0.7,-1.6) -- (1.951834,0.436286);
       \end{scope}
       \begin{scope}[xshift=0.0cm,yshift=0.0cm]
       \draw[ultra thick] (0,0) circle (2);
       \end{scope}
       \begin{scope}[xshift=3.903668cm,yshift=0.0cm]
       \draw[thick,decorate,decoration={snake,amplitude=.3mm,segment length=2mm,pre length=1.0mm, post length=1.0mm}] (-1.951834,0.4362868)  -- (-0.61885,1.90211); 
       \draw[thick,decorate,decoration={snake,amplitude=.3mm,segment length=2mm,pre length=1.0mm, post length=1.0mm}] (-0.61885,1.90211) -- (1.721484,1.01808) ; 
       \draw[thick,decorate,decoration={snake,amplitude=.3mm,segment length=2mm,pre length=1.0mm, post length=1.0mm}] (1.721484,1.01808) -- (0.7,-1.6); 
       \draw[thick,decorate,decoration={snake,amplitude=.3mm,segment length=2mm,pre length=1.0mm, post length=1.0mm}] (0.7,-1.6)-- (-0.85156,-1.809654); 
       \draw[thick,decorate,decoration={snake,amplitude=.3mm,segment length=2mm,pre length=1.0mm, post length=1.0mm}] (-0.85156,-1.809654) -- (-1.2,0.2); 
       \draw[thick,decorate,decoration={snake,amplitude=.3mm,segment length=2mm,pre length=1.0mm, post length=1.0mm}] (-1.2,0.2)-- (0.4,0.15); 
       \end{scope}
       \begin{scope}[xshift=3.903668cm,yshift=0.0cm]
       \draw[ultra thick] (0,0) circle (2);
       \end{scope}
       \begin{scope}[xshift=1.951834cm,yshift=0.0cm]
       \draw[ultra thick] (0,0) circle (3.951834);
       \end{scope}
  \end{tikzpicture}
  \end{center}
  \caption{ Merging bounding circles for two sub-chains to form the
  bounding circle for the concatenated chain in $\R^2$.\label{fig:circle}}
\end{figure}
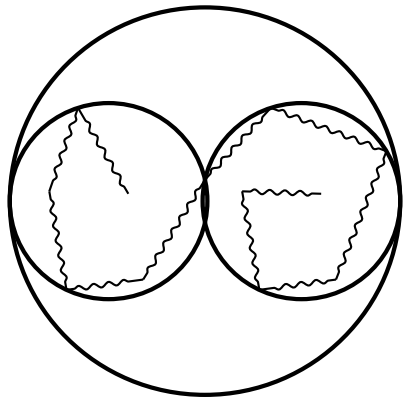

One of the operations that must be performed with bounding
volumes~\cite{Clisby2010Efficientimplementationpivot} is the merge
operation, which involves combining two bounding volumes (which contain
sub-chains) to create a bounding volume that contains both of the
original bounding volumes (and hence contains both sub-chains).  In
contrast to the situation for lattice models, the bounding volumes which
result from the merge operation do not necessarily form a snug fit for
the polymer sub-chains.  This is illustrated in Fig.~\ref{fig:circle}
for an example in $\R^2$ where the snugly fitting bounding circles for
two sub-chains are merged together so that they contain the concatenated
walk. The concatenated walk \emph{does not} touch the boundary of the
larger circle.

\emph{A priori}, we had no expectation about whether the lack of snug
fit for the bounding volumes would prove to be a significant problem. We
considered it possible that the error from the fit would grow rapidly as
one moved up the SAW-tree, and this would have worsened the performance
of the intersection testing algorithm.
But, we found that in fact this was not a problem at all. We estimated
the mean ratio of the diameter of the bounding volume to the square root
of the mean
value of the squared end-to-end distance $\langle R_E^2 \rangle^{1/2}$.
We found that as the length of the chains increased the ratio was
approaching a constant for both $\R^2$ and $\R^3$, indicating that the
error was becoming saturated. For chain lengths of $N=10^6$ this ratio
was only 1.45 for $\R^2$, and 1.71 for $\R^3$.
Thus, in the average case this suggests that the lack of a snug fit only
results in a constant factor error in the diameter of the bounding
volume for the off-lattice implementation.
This means that the behaviour of the lattice and off-lattice
implementations should be essentially the same, up to a constant factor.

\begin{figure}[htb]
\begin{center}
\includegraphics[width=0.5\paperwidth]{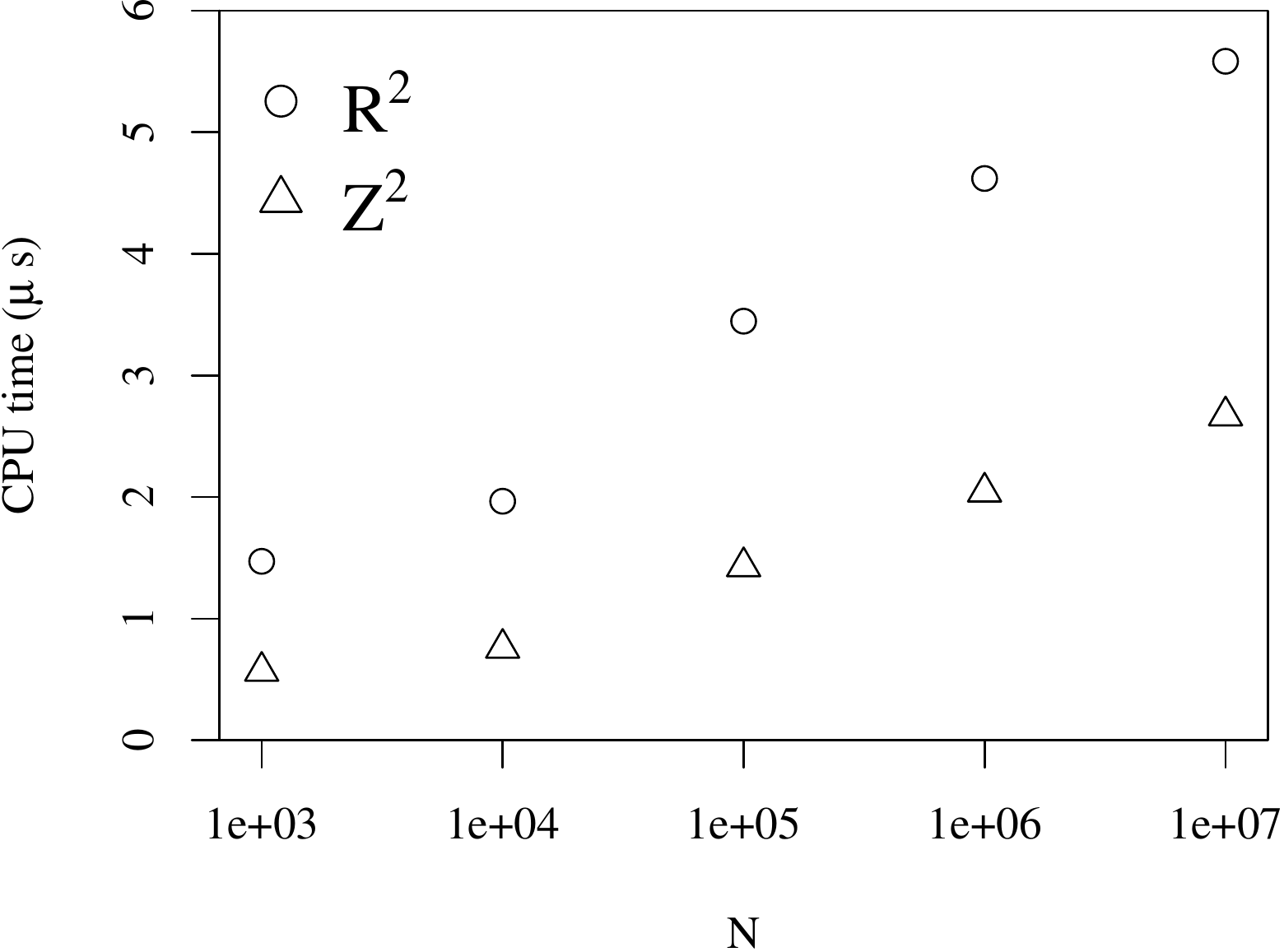}
\end{center}
\caption{CPU time per pivot move attempt for the bead-necklace model in $\R^2$,
in comparison to SAWs in $\Z^2$, plotted against the number of monomers
$N$.\label{fig:cpud2}}
\end{figure}

\begin{figure}[htb]
\begin{center}
\includegraphics[width=0.5\paperwidth]{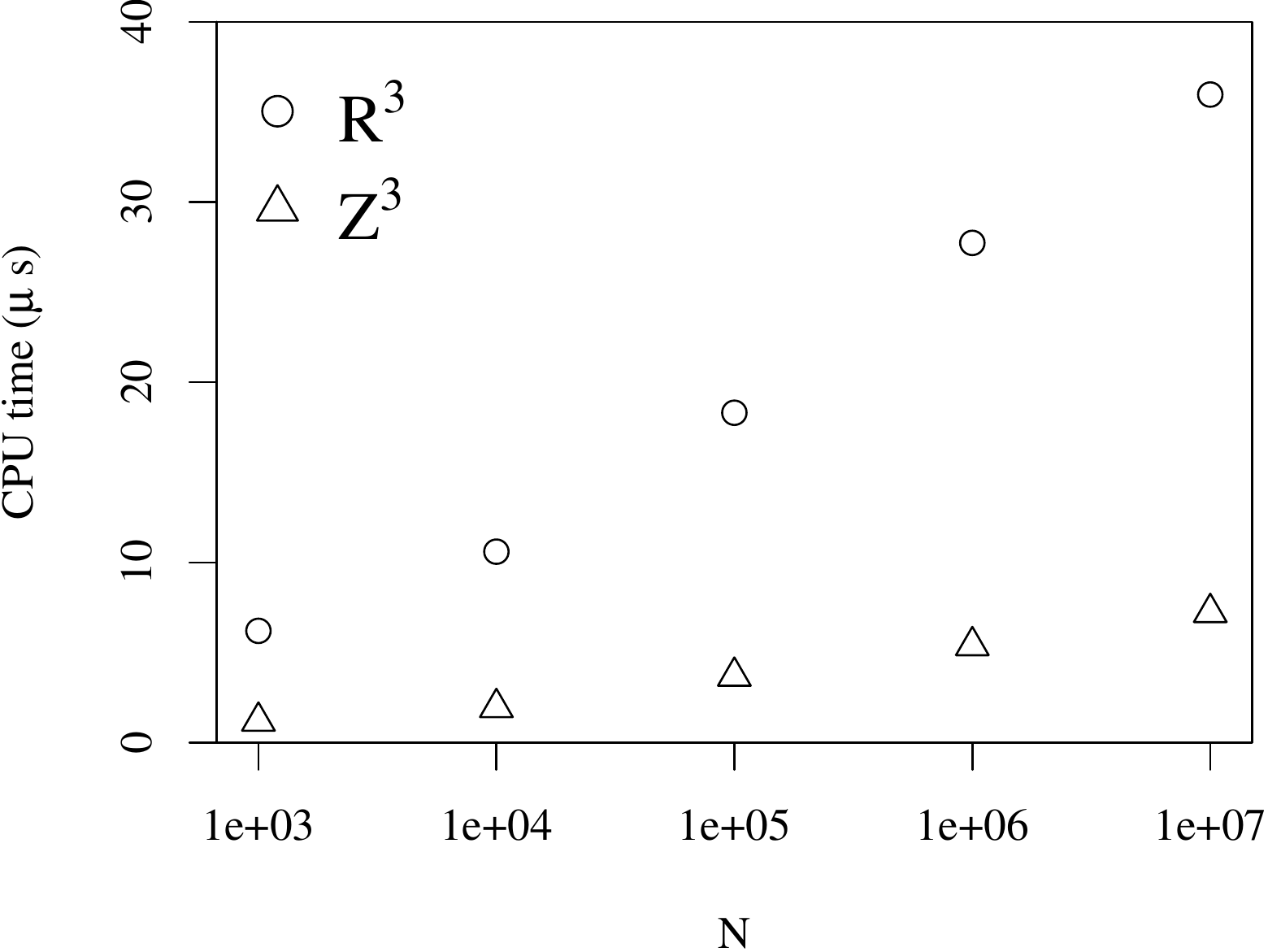}
\end{center}
\caption{CPU time per pivot move attempt for the bead-necklace model in $\R^3$,
in comparison to SAWs in $\Z^3$, plotted against the number of monomers
$N$.\label{fig:cpud3}}
\end{figure}

We evaluated the mean CPU time per pivot move for a range of polymer
lengths, for lattice and off-lattice SAW-tree implementations in
two and three dimensions on Dell
PowerEdge FC630 machines with Intel Xeon E5-2680 CPUs, and plot the
results of these computer experiments in Figs~\ref{fig:cpud2} and
\ref{fig:cpud3}.

We found that the time per pivot move attempt was somewhat worse for the
off-lattice implementation as compared to the lattice implementation,
which was to be expected due to the increased number of operations required
for computations
involving the symmetry elements and coordinate vectors. But, in absolute
terms the performance is still impressive, and for polymers with
$10^7$ monomers pivot attempts are performed in mean CPU time of less
than 6$\mu$s for $\R^2$, and in less than 40$\mu$s for $\R^3$.
We clearly
observe $O(\log N)$ behaviour in each case, which is strong evidence
that the off-lattice implementation behaves in fundamentally the same
way as the original lattice implementation of the SAW-tree.

\section{Parallel implementation of the pivot algorithm}
\label{sec:parallel}

The SAW-tree implementation of the pivot
algorithm~\cite{Clisby2010Efficientimplementationpivot} is remarkably
efficient, but it suffers from one significant drawback: the
intersection testing and SAW-tree update procedures are inherently
serial operations. 
This makes it difficult to take advantage of additional cores to improve
the rate at which polymer configurations are sampled.
To some extent this issue is obviated by the fact that for number of
monomers $N$ up to the order of tens of millions or even 100 million it
is possible to run simulations in parallel on multicore machines, and
still obtain results in a reasonable clock time.

But, in the regime where a large amount of memory is needed for truly
large $N$, of the order of $10^8-10^9$, on the Dell PowerEdge
FC630 machines with Intel Xeon E5-2680 CPUs on which computer
experiments are being run this prevents all cores being
simultaneously used due to memory constraints\footnote{There are 24 cores,
and total memory available is 128GB.}. Under these circumstances most
cores must be left idle while data is being collected.

Here we will briefly sketch a method to improve the sampling rate by
utilising additional cores in exactly this difficult regime.

The key insight is that as the number of monomers increases, the
probability of a pivot move being successful decays as a power law of
the form $N^{-p}$, with $p \approx 0.19$ for $\Z^2$, and $p \approx
0.11$ for $\Z^3$. For $N = 10^9$ on $\Z^2$, the probability of a pivot
move being successful is 0.019, which means that on average roughly
50 unsuccessful pivot attempts are made for each success.

Given that most proposed pivot moves in this regime fail, and so do not
result in any update being made for the self-avoiding walk, it is
possible to perform many pivot attempts in parallel without this effort
being wasted.

For example, imagine that we are sampling SAWs of $10^9$ steps via the
pivot algorithm, and we may test for success or failure of up to ten pivot
moves simultaneously. Note that a move consists of a proposed monomer
location to act as the centre of the pivot move, and a proposed symmetry
operation. Suppose for the first batch of ten proposed moves
$\{M_1, M_2, \cdots, M_{10}\}$,
that each of these moves were unsuccessful. Then, we can move on to
another batch, and none of the work performed by any of the threads was
wasted. Suppose for the second batch $\{M_{11}, M_{12}, \cdots,
M_{20}\}$ that the first 6 moves $M_{11},\cdots,M_{16}$ are
unsuccessful, but $M_{17}$ is successful. Then we need to perform the
update associated with the move $M_{17}$ which must happen as a serial
operation performed by a single thread. It does not matter whether
$M_{18}, M_{19}, M_{20}$ were successful or not: these tests will need
to be performed again in case the update has altered the result of the
test. The next batch will then consist of ten proposed moves
$\{M_{18}, M_{19}, \cdots, M_{27}\}$.

The tests for success or failure will occur for each thread regardless
of the outcome of the tests performed by other threads. But, provided
the probability of multiple successful moves occurring in a batch is
low, then most of this work will not be wasted.
The lower the probability of success, the greater the potential for
speed up to occur by exploiting parallelism.

We have implemented this idea in a prototype C program with OpenMP being used for
managing the parallel pivot attempts. The SAW-tree is held in shared
memory where all threads can access it for performing intersection
tests. When a pivot move is found to be successful, then the update is
performed by a single thread while all other threads remain idle.

We performed computer experiments to test this implementation on the
aforementioned FC630 machines for SAWs of various lengths on the square lattice. We utilised 24 threads, with batches (or
chunks) of 48 pivot attempts which meant that each thread made two
attempted pivot moves on average. We collated the calendar time per
pivot attempt in $\mu$s in Table~\ref{tab:parallel}. 
The value $t_1$ is the mean CPU time for a single thread, while $t_{24}$
is the mean CPU time for the 24 threads running in parallel.
We see that as $N$ increases the probability of a move being successful
decreases, and the relative performance of the parallel implementation
to the serial implementation improves.
For $N = 10^9$ there is
roughly a four-fold improvement in performance.

Although it is suitable as a proof-of-concept, the implementation
developed thus far is only a prototype, and more work remains to be done
to improve its performance. 
In particular, it should be possible to re-use some
information from intersection tests even if these moves are scheduled to
occur after a move that is found to be successful. For example, if a
move is found to cause a self-intersection between monomers labelled $l$
and $m$ along the chain, then if the prior succesful move involved a pivot
site outside of the interval $l$ to $m$ then this would not have any
effect on the self-intersection. 
Nonetheless, even in its current state the
performance gain is sufficient to make it worthwhile for use in the large
$N$, memory-limited regime.

\begin{table}[htb]
\caption{Performance of the prototype parallel implementation of the pivot
algorithm for $\Z^2$.\label{tab:parallel}}
\begin{center}
\begin{tabular}{rlllll}
\hline
$N$   & $\Pr(\text{success})$  & $1/\Pr(\text{success})$& $t_1$ ($\mu$s) & $t_{24}$ ($\mu$s) & $t_1/t_{24}$ \\
\hline
\tstrut $10^6$& 0.068& 15   & 1.58 & 1.35    & 1.17         \\
$10^7$& 0.044& 23   & 2.31 & 1.07    & 2.16         \\
$10^8$& 0.029& 34   & 2.90 & 0.903   & 3.21         \\
\bstrut $10^9$& 0.019& 53   & 3.16 & 0.805   & 3.93         \\
\hline
\end{tabular}
\end{center}
\end{table}







\section{Discussion and conclusion}
\label{sec:conclusion}

Schnabel and
Janke~\cite{Schnabel2019}
have very recently implemented a binary tree data structure which is
similar to the SAW-tree for the bead-necklace model, as well as a model
for which the Lennard-Jones interaction is implemented. The
implementation for the bead-necklace model appears to have roughly the
same computational efficiency as the implementation sketched here. The
efficient implementation for the Lennard-Jones polymer model is very
interesting, and a significant advance on the state of the art. It will
be interesting to see if further progress in this direction can be made,
for example in the evaluation of Coulomb interactions
which would be necessary for efficient simulation of polyelectrolytes.

Full details for the off-lattice SAW-tree implementation of the pivot
algorithm will be presented elsewhere in future.

More work needs to be done to test and improve the implementation of the
parallel version of the pivot algorithm.
In future, the parallel implementation of the pivot algorithm will allow for
improved simulations of very long SAWs on the square lattice. The method
will result in significant speed-ups for SAWs with hundreds of millions
or even one billion steps, especially for the square lattice.

\section*{References}


\begin{thebibliography}{1}
\expandafter\ifx\csname url\endcsname\relax
  \def\url#1{{\tt #1}}\fi
\expandafter\ifx\csname urlprefix\endcsname\relax\def\urlprefix{URL }\fi
\providecommand{\eprint}[2][]{\url{#2}}

\bibitem{Lal1969MonteCarlocomputer}
Lal M 1969 {\em Mol. Phys.\/} {\bf 17} 57--64

\bibitem{Madras1988PivotAlgorithmHighly}
Madras N and Sokal A~D 1988 {\em J. Stat. Phys.\/} {\bf 50} 109--186

\bibitem{Kennedy2002fasterimplementationpivot}
Kennedy T 2002 {\em J. Stat. Phys.\/} {\bf 106} 407--429

\bibitem{Clisby2010AccurateEstimateCritical}
Clisby N 2010 {\em Phys. Rev. Lett.\/} {\bf 104} 055702

\bibitem{Clisby2010Efficientimplementationpivot}
Clisby N 2010 {\em J. Stat. Phys.\/} {\bf 140} 349--392

\bibitem{Clisby2018MonteCarlo4dSAWs}
Clisby N 2018 {\em J. Stat. Phys.\/} {\bf 172} 477--492

\bibitem{Stewart1980EfficientGenerationOfRandomOrthogonalMatrices}
Stewart G~W 1980 {\em SIAM Journal on Numerical Analysis\/} {\bf 17} 403--409

\bibitem{Plunkett2016OffLatticeSAWPivotAlgorithmVariant}
Plunkett L and Chapman K 2016 {\em J. Phys. A: Math. Theor.\/} {\bf 49}
  135203 

\bibitem{Schnabel2019}
Schnabel S and Janke W  (\textit{Preprint} \eprint{arXiv:1904.11191})

\end{thebibliography}

\providecommand{\newblock}{}

\section*{Acknowledgements}
  Thanks to Stefan Schnabel for communicating results regarding an
  alternative efficient off-lattice implementation of the pivot
  algorithm prior to
  publication.
  N.C. gratefully acknowledges support from the Australian Research Council under
  the Future Fellowship scheme (project number FT130100972). 
\end{document}